\documentclass[a4paper,fleqn,usenatbib]{mnras}

\usepackage{newtxtext,newtxmath}

\usepackage[T1]{fontenc}
\usepackage{ae,aecompl}

\usepackage{graphicx}	
\usepackage{amsmath}	
\usepackage{amssymb}	
\usepackage{ulem}
\usepackage{gensymb}
\usepackage{hyperref}
\usepackage{threeparttable}
\usepackage{float}
\usepackage{enumitem}

\title[Cluster galaxy kinematics via VDPs]{The Kinematics of Cluster Galaxies via Velocity Dispersion Profiles}

\author[L. E. Bilton and K. A. Pimbblet]{
Lawrence E. Bilton$^{1}$\thanks{E-mail: l.bilton@2016.hull.ac.uk}
and Kevin A. Pimbblet,$^{1}$
\\
$^{1}$E.A. Milne Centre for Astrophysics, The University of Hull, Cottingham Road, Kingston upon Hull, HU6 7RX, UK
}

\date{Accepted XXX. Received YYY; in original form ZZZ}

\pubyear{2018}

\begin{document}
\label{firstpage}
\pagerange{\pageref{firstpage}--\pageref{lastpage}}
\maketitle

\begin{abstract}
	We present an analysis of the kinematics of a sample of 14 galaxy clusters via velocity dispersion profiles (VDPs), compiled using cluster parameters defined within the X-Ray Galaxy Clusters Database (BAX) cross-matched with data from the Sloan Digital Sky Survey (SDSS). We determine the presence of substructure in the clusters from the sample as a proxy for recent core mergers, resulting in 4 merging and 10 non-merging clusters to allow for comparison between their respective dynamical states. We create VDPs for our samples and divide them by mass, colour and morphology to assess how their kinematics respond to the environment. To improve the signal-to-noise ratio our galaxy clusters are normalised and co-added to a projected cluster radius at $0.0 - 2.5$ $r_{200}$. We find merging cluster environments possess an abundance of a kinematically-active (rising VDP) mix of red and blue elliptical galaxies, which is indicative of infalling substructures responsible for pre-processing galaxies. Comparatively, in non-merging cluster environments galaxies generally decline in kinematic activity as a function of increasing radius, with bluer galaxies possessing the highest velocities, likely as a result of fast infalling field galaxies. However, the variance in kinematic activity between blue and red cluster galaxies across merging and non-merging cluster environments suggests galaxies exhibit differing modes of galaxy accretion onto a cluster potential as a function of the presence of a core merger.
\end{abstract}





\section{Introduction}
\label{sec:intro}

Galaxies are known to follow a morphology-density relation, which is pronounced in clusters of galaxies (\citealt{Oemler1974}; \citealt{Dressler1980}; \citealt{Smith2005}). 
Late-type galaxies are found to dominate at large radii from a galaxy cluster centre, predominantly within the field population. Conversely, early-type galaxies are found to pervade the denser regions at smaller radii, well within galaxy clusters. 
There are further observable environmental side-effects that follow similar patterns, such as the apparent bimodality of the colour-density relation \citep{Hogg2003,Hogg2004}, where denser regions are populated with quenched, red and elliptical galaxies. 
Contrarily star-forming, blue and spiral morphologies are found out towards the field population (e.g. \citealt{Lewis2002}; \citealt{Gomez2003}; \citealt{Bamford2009}). 
 
Galaxy clusters are consequently an epicentre for environmental interactions. 
The comparative accretion histories of cluster galaxies between galaxy clusters and the field population can be determined as a function of their environment, indicated by their membership's morphology, colour and star-formation assuming a fixed stellar mass (e.g. \citealt{Postman1984}; \citealt{Hogg2004}; \citealt{Linden2010}). 
The evolutionary transformation of cluster galaxies could have transpired prior to a galaxy's accretion onto a cluster's potential, since the field population's morphologies, colours and rate of star-formation are mixed (e.g. \citealt{Kauffmann2004}; \citealt{Blanton2005}). 
Or, it is possible that the harassment and accretion of a galaxy by a cluster leads onto a transformation of blue to red; star-forming to non-star-forming; spiral to elliptical \citep{Moore1996}. 
This metamorphosis during the infall of a galaxy into a cluster is considered to be the result of an increased probability of tidal galaxy-galaxy interaction mechanisms, or, even the tidal field of the cluster itself. 
The former being more likely to give rise to the stripping of material, and distortion of a galaxy's structure \citep{Moore1999}. 
Further observations ostensibly show the shifting of morphologies from late-type to early-type are chiefly to be the result of mergers between two galaxies (e.g. \citealt{Owers2012}). 
 
The volume between cluster galaxies contains a sea of hot diffuse gas that represents an intracluster medium (ICM), another form of environmental interaction. 
An infalling galaxy approaching a cluster centre at higher velocities relative to the ICM will experience ram pressure stripping  (\citealt{Gunn1972}; \citealt{Abadi1999}; \citealt{Quilis2000}; \citealt{Sheen2017}). 
The disc of cold gas surrounding an infalling galaxy will be stripped away over small timescales, however, as the ICM density increases during infall so do the time scales of this process \citep{Roediger2007}. 
The result of this process retards rates of star-formation to where the infalling galaxy will be quenched completely. 
The diffuse nature of any hot gas haloes surrounding infalling galaxies lends to their increased likelihood of being ejected from the galaxy's potential. 
Therefore, the removal of any surrounding haloes of hot gas around an infalling galaxy will inhibit the replenishment of their cold gas reservoirs through radiative cooling, slowly strangling galaxy star-formation, with any remaining cold gas being exhausted  \citep{Larson1980}. 
Ram pressure stripping has been found to be prevalent in the dense cores of clusters through observations of tails with \ion{H}{I} and H$\alpha$ emission lines that are associated with a parent galaxy (e.g. \citealt{Gavazzi2001}; \citealt{Cortese2007}). 
 
With galaxy cluster environments hosting extended ICM haloes that interact significantly with field and infalling galaxies, consideration of a cluster's size is therefore needed in order to understand where the boundary between these environments lie.
One common definition of a cluster's size is the virial radius, commonly approximated as $R_{vir}\sim r_{200}$. 
$r_{200}$ represents the radial point at which the average density is $\sim200$ times the critical density (e.g. \citealt{Linden2010}; \citealt{Pimbblet2012}; \citealt{Bahe2013}; \citealt{Pimbblet2014}). 
However, the proposed splashback radius may represent a more physical boundary, extending farther than $r_{200}$ (e.g. \citealt{More2015}; \citealt{More2016}; \citealt{Baxter2017}). 
The splashback radius represents the first apoapsis of an observed accreted galaxy that has already passed through its first periapsis or turnaround (\citealt{Sanchis2004}; \citealt{Pimbblet2006}). 
Despite both of these definitions for a potential cluster boundary, they do not extend to the radii observed with harassed galaxies infalling to the cluster centre; colour-densities and effects on star-formation can continue beyond these defined boundaries (e.g. \citealt{Balogh1999}; \citealt{Haines2009}; \citealt{Linden2010}; \citealt{Haines2015}). 
A plethora of observations and simulations appear to indicate that there is a natural fluidity between the local cluster environment and the field population of late-type star-forming galaxies. 
Such simulations have shown the entire cluster boundary to expand even grander scales with ICM haloes extending out to radii of $\sim10$ Mpc from the cluster centre \citep{Frenk1999}.

The existence of these large-scale structures therefore indicates the presence of smaller scale clumping of galaxies; more layers of substructure within galaxy clusters are expected (\citealt{Dressler1988}). 
It is more likely that any accreted galaxies from the field population will undergo \textquoteleft pre-processing' into smaller galaxy groups that help form the substructure within a cluster (e.g. \citealt{Berrier2009}; \citealt{Bahe2013}), inducing evolutionary changes prior to traditional cluster galaxy infall and accretion. 
In the simulation work of \cite{Haines2015} it is found that star-forming galaxies are unexpectedly quenched at large radii from the cluster centre, models can only account for this if the galaxies have undergone pre-processing into a substructure prior to any further interaction. 
There is an alternative variant of pre-processing in rarer cluster-cluster merger events, the most famous example of such an event is the Bullet Cluster \citep{Tucker1998}. 
X-ray observations of the Bullet Cluster show a smaller sub-cluster of galaxies colliding with a larger cluster, thereby ram-pressure stripping causing the removal of the surrounding hot gas \citep{Markevitch2002}. 
Other \textquoteleft bullet-like' events are shown to effect the local galactic environment in equivalent ways (e.g. \citealt{Owers2011}; \citealt{Owers2012}).

This leads on to potential ways to make a comparison between these different environments via their varying dynamical states. 
We can therefore probe the variation in cluster environments via analysis of the cluster kinematics as a function of radius with Velocity Dispersion Profiles. 
VDPs represent how the radial velocity dispersions vary from the dense area of accreted early-type galaxies within $r_{200}$, out to sparser star-forming late-types on their infall journey to the centre (see \citealt{Hou2009,Hou2012}). 
It is therefore possible to test how the shape of a VDP is affected by binning a profile based on different cluster galaxy properties. 
As an example, \cite{Pimbblet2012} splits the VDP of Abell 1691 into individual high and low mass profiles. 
It is found that there is a large disparity in the velocities between the high and low mass samples, \cite{Pimbblet2012} argues the large high mass sample velocities could be due to the presence of substructure, or recent arrivals to the system. 
The shape of the VDP could, however, be affected by any evolutionary change due to the cluster environment. 

In this work, we aim to test how the average cluster VDP's shape can be altered as a function of radius, parameterised by its member's different evolutionary stages through proxies of varying masses, colours and morphologies, in order to explore the varying dynamics between merging, dynamically active and non-merging, relaxed environments. 
We therefore present galaxy data taken from the Sloan Digital Sky Survey (SDSS) to form a membership from a defined cluster sample determined from an X-ray catalogue. 
Details on how the data was acquired can be found in section~\ref{sec:data}. 
Details on the derivation and production of the VDPs can be found in section~\ref{sec:vdp}. 
A discussion of the data, results and their consequences are outlined in section~\ref{sec:disc}, followed by a summary of our conclusions in section~\ref{sec:conc}.

Throughout the work presented here we assume a $\Lambda$CDM model of cosmology with $\Omega_{M} = 0.3$, $\Omega_{\Lambda} = 0.7$, $H_{0} =$ $100h$ km s$^{-1}$ Mpc$^{-1}$, where $h = 0.7$.

\section{The Data}
\label{sec:data}

We define a sample of galaxy clusters using the X-Ray Galaxy Clusters Database (BAX, \citealt{Sadat2004}), a comprehensive catalogue of X-ray emitting clusters from multiple literary sources. For each galaxy cluster we obtain members from SDSS Data Release 8 (DR8, \citealt{Aihara2011}) with complementary data from MPA-JHU Value Added Catalogue (\citealt{Kauffmann2003}; \citealt{Brinchmann2004}; \citealt{Tremonti2004}). We use data from Galaxy Zoo 2 (GZ2, \citealt{Willett2013}; \citealt{Hart2016}) to provide morphological information on member galaxies.

\subsection{Defining the Cluster Sample and their Membership}
\label{sec:dcm}

To select our cluster sample, we adopt an X-ray luminosity range of $3 < L_{X} < 30\times10^{44}$ ergs s$^{-1}$. 
These limits ensure we are selecting the most massive clusters from the BAX catalogue across a range of dynamically relaxed and perturbed states. 
We impose a redshift range of $0.0 < z < 0.1$, which serves to help make the final sample of galaxies making each cluster complete.
The imposed limits with BAX output a base sample size of 68 clusters.

\begin{figure*}
	\includegraphics[width=0.85\paperwidth]{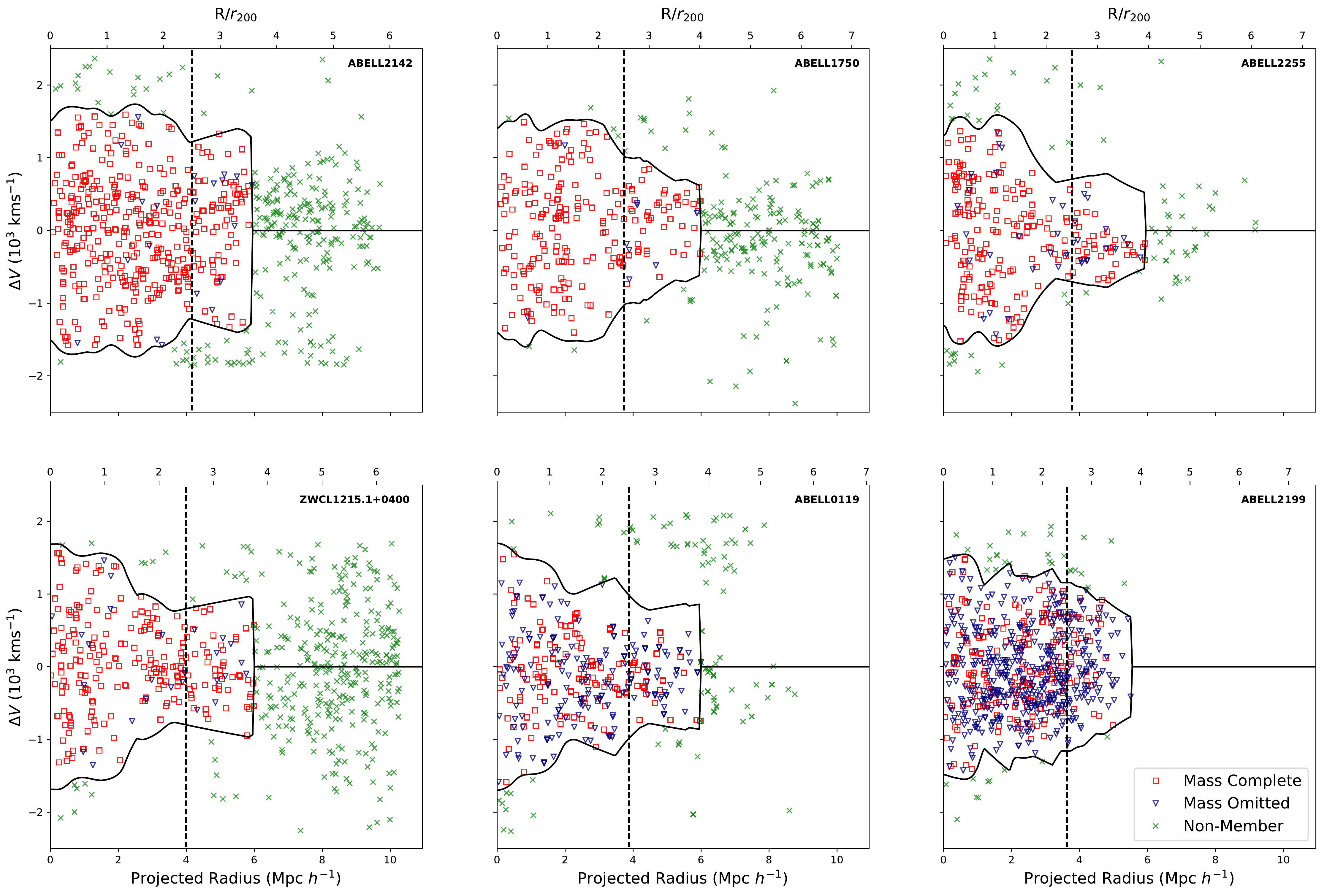}
	\caption{A selection of example surface caustics (the black curves) from the final merging cluster sample (top row) and non-merging cluster sample (bottom row). Where the red squares represent the galaxies that make a complete sample at log$_{10}$(M$_{\ast}$) $\geq10.1$, with the blue triangles representing omitted galaxies that are at log$_{10}$(M$_{\ast}$) $<10.1$. Galaxies that lie within the surface caustics are considered to be cluster members. Here the radial velocity ($\Delta V$) with respect to the cluster's mean recession velocity is plotted against the projected radius in units of Mpc $h^{-1}$ and $R/r_{200}$. The black dashed vertical lines represent the 2.5 $R/r_{200}$ radial cut of each cluster. Only galaxies of $\leq$2.5 $R/r_{200}$ within the caustics are used in the production of the stacked VDPs.}
	\label{fig:caus}
\end{figure*}

For each of the clusters in the sample a $10$ Mpc $h^{-1}$ upper radial limit of DR8 galaxies is applied from the BAX defined centre to the appropriate scales, using the flat cosmology prescribed in section \ref{sec:intro} \citep{Wright2006}. 
Each candidate cluster have their global means ($\overline{cz}_{glob}$) and velocity dispersions ($\sigma_{glob}$) calculated for galaxies $\leq1.5$ Mpc $h^{-1}$, the latter are determined by the square root of the biweight midvariance (see \citealt{Beers1990}). 
Due to a willingness to observe the effect infall galaxies have on velocity dispersion profiles beyond $r_{200}$, a constant line boundary applied in velocity space is not ideal to distinguish an infaller from the field, since a cluster's potential varies with increasing $R$ from the centre. 
Using the mass estimation method of caustics (\citealt{Diaferio1997}; \citealt{Diaferio1999}), we produce surface caustics with velocity limits of $\Delta V=\pm$ 1500 kms$^{-1}$ and a radial limit of $R \leq$ 10 Mpc $h^{-1}$, where $\Delta V = c[(z_{gal}-z_{clu})/(1+z_{clu})]$. 
The surface caustics help determine the final membership that considers the varying potential as a function of $R$ (\citealt{Gifford2013}; \citealt{Gifford2013a}). 
The resultant caustic mass profiles allow for an estimation of $r_{200}$ with the application of a varying enclosed density profile, $\rho(r)=3M(r)/4\pi r^{3}$, until $\rho(r)=200\rho_{c}$, where $\rho_{c}$ is the critical density of the universe for our flat cosmology.
An example of these surface caustics are shown in Figure \ref{fig:caus} and are discussed in Section \ref{sec:psc}.

The final values for $\sigma_{glob}$, $\sigma_{r_{200}}$ for galaxies $\leq$ $r_{200}$ and $\overline{cz}_{glob}$ are determined.  
The uncertainties for these parameters are calculated following the methodology of \cite{Danese1980}. 
In order to maximise the number of DR8 galaxies per cluster while maintaining a mass-complete sample across our redshift range, we impose a stellar mass limit of log$_{10}$(M$_{\ast}$) $\geq10.1$. 
Candidate clusters are then cross-checked with the \cite{Einasto2001} catalogue of superclusters to help eliminate those structures that overlap with one another.
A final check we employ before a cluster is added to the final master sample is to test if the cluster is sufficiently
rich in its membership of cluster galaxies. 
We define the richness limit here as clusters with >50 galaxies at $\leq r_{200}$, any clusters not meeting this requirement are ignored. 
This leads to a resultant sample size of 14 galaxy clusters.

\begin{figure}
	\includegraphics[width=\columnwidth]{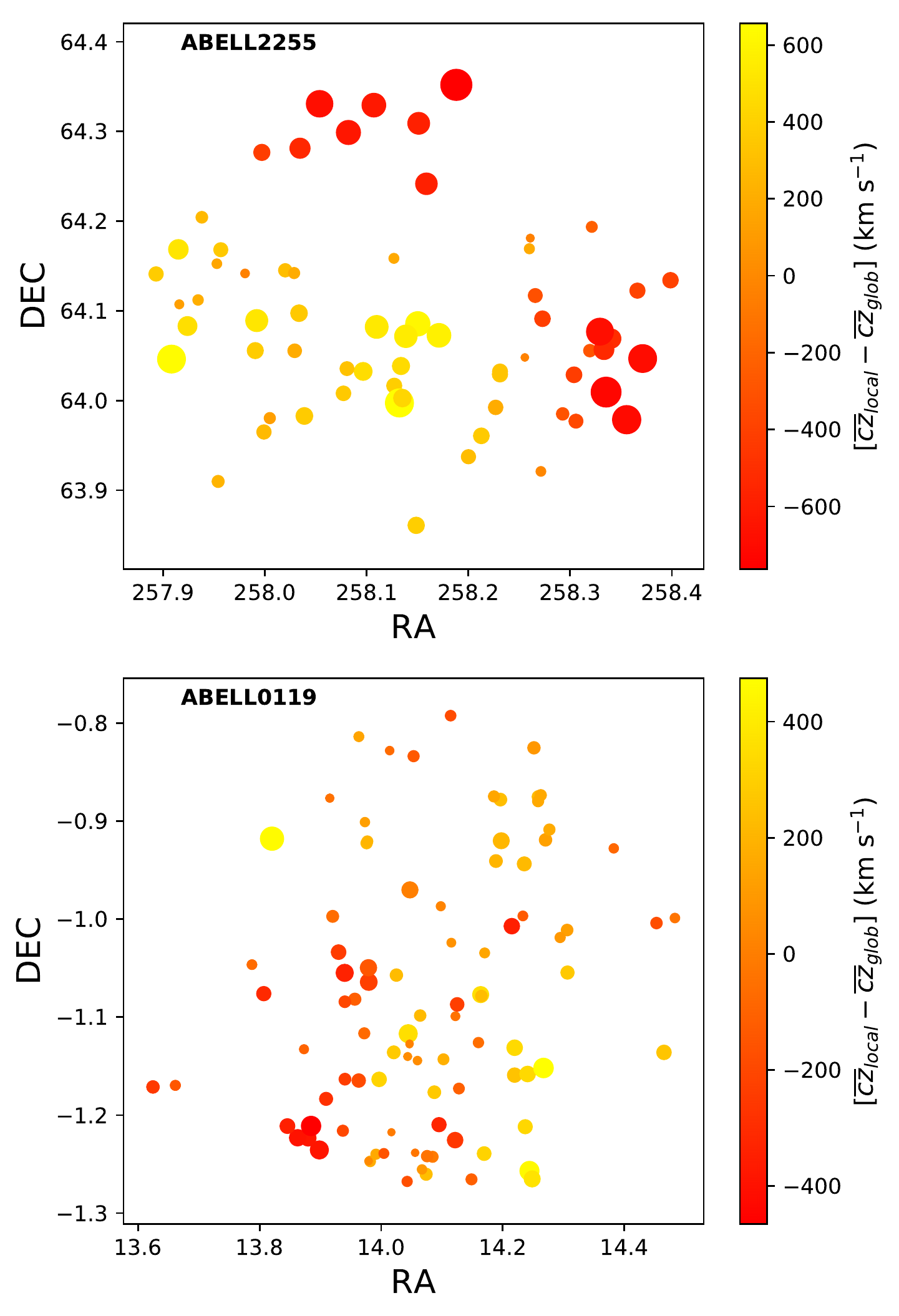}
	\caption{Example bubble plots from the $\Delta$ test, where the total area of each circle is proportional to the deviation $e^{\delta_{i}}$, and the colours representing varying radial velocity differences $[\overline{cz}_{local} - \overline{cz}_{glob}]$. Cluster Abell 2255 (top) shows significant sub-clustering with a greater number of substantial deviations from the global values, as demonstrated by the overlapping larger area circles with large radial velocities. Cluster Abell 0119 (bottom) in comparison demonstrates weak sub-clustering, with fewer numbers of significantly strong deviations from the global values.}
	\label{fig:bub}
\end{figure}

\subsection{Merging Cluster Sample}
\label{sec:merg}
For the thesis of this work, we create subset samples of merging and non-merging galaxy cluster systems in order to compare how their respective environments affect the kinematics of their members. 
To determine whether or not a cluster is merging we employ the $\Delta$ test of substructure devised by \cite{Dressler1988} on galaxies $\leq 1.5$ Mpc $h^{-1}$ from the BAX defined cluster centre. 
The $\Delta$ test methodology takes the local mean radial velocity ($\overline{cz}_{local}$) and local standard deviation of the radial velocity ($\sigma_{local}$) of a galaxy and its $N_{nn}=\sqrt{N_{glob}}$ nearest neighbours, where $N_{glob}$ is the number of galaxies $<1.5$ Mpc $h^{-1}$. 
These are then compared to the global values of the cluster they are the members of, as shown in equation \ref{eq:ds}. 

\begin{equation}
	\delta^{2}_{i} = \left(\frac{N_{nn}+1}{\sigma_{glob}^{2}}\right) [(\overline{cz}_{local} - \overline{cz}_{glob})^{2} + (\sigma_{local} - \sigma_{glob})^{2}],
	\label{eq:ds}
\end{equation}

\noindent
where $\delta$ measures the deviation in the small region around the galaxy compared to the global cluster values at $\leq1.5$ Mpc $h^{-1}$. 
This process is iterated through each galaxy to produce the sum $\Delta = \sum_{i}\delta_{i}$. 
\cite{Pinkney1996} has shown the $\Delta$ test to be the most sensitive for indicating the presence of substructure, demonstrating a $\geq$ 99\% significance in determining its occupancy. 
Therefore, a cluster will be classified as merging when substructure is detected at $P(\Delta)\leq0.01$. 
All clusters that reject the null hypothesis are added to the subset merging cluster sample. 
We discuss some of the caveats of this approach in section \ref{sec:DT}.
The resultant merging subset sample contains 4 galaxy clusters, detailed in Table \ref{tab:sub}, leaving the non-merging subset outweighing the mergers with 10 galaxy clusters, detailed in Table \ref{tab:nsub}. 
Example bubble plots of a merging and non-merging cluster from both samples are shown in Figure \ref{fig:bub}, where the area of each circle is proportional to $e^{\delta_{i}}$, indicating the level of sub-structuring through the magnitude of deviations from the global values.

\begin{table*}
	\centering
	\caption{The mass complete merging cluster subset sample. The J2000 coordinates and X-ray luminosity values are taken from BAX. $\sigma_{r_{200}}$ is determined from a biweight estimator, as noted by Beers et al. (1990). The uncertainties for the mean recession velocities and velocity dispersions are calculated following the method by Danese et al. (1980). The $\sigma_{ref}$ values are reference velocity dispersions from the literature. The $P(\Delta)$ values testing for substructure follow the methods of Dressler \& Shectman (1988) with equation \ref{eq:ds}, those values that are $\ll$0.01 strongly reject the null hypothesis and have values smaller than to three decimal places.}
	\label{tab:sub}
	\begin{tabular*}{\textwidth}{l@{\extracolsep{\fill}}cccccccr} 
		\hline
		Name & RA  & Dec & $L_{x}$ & $N_{r_{200}}$ & $\overline{cz}_{glob}$ & $\sigma_{r_{200}}$ & $\sigma_{ref}$ & $P(\Delta)$\\
				 & (J2000) & (J2000) & ($\times 10^{44}$ erg s$^{-1}$) & & (km s$^{-1}$) & (km s$^{-1}$) & (km s$^{-1}$) &  \\
		\hline
		Abell 426 & 03 19 47.20 & +41 30 47 & 15.34$^{a}$ & 97 & 5155$\pm 59$ & 827$^{+40}_{-47}$ & 1324$^{1}$ & 0.010 \\
		
		Abell 1750 & 13 30 49.94 & -01 52 22 & 5.98$^{b}$ & 72 & 25614$\pm 92$ & 782$^{+56}_{-72}$ & 657$^{2}$ & $\ll$0.01 \\
		
		Abell 2142 & 15 58 20.00 & +27 14 00 & 21.24$^{a}$ & 132 & 26882$\pm 84$ & 816$^{+52}_{-63}$ & 1193$^{8}$ & 0.005 \\
		
        Abell 2255 & 17 12 31.05 & +64 05 33 & 5.54$^{a}$ & 72 & 24075$\pm 98$ & 788$^{+60}_{-79}$ & 1009$^{4}$ & $\ll$0.01 \\
        
		\hline
	\end{tabular*}
	\begin{tablenotes}
		\item $^{1}$ \cite{Struble1999} \hfill $^{a}$ \cite{Reiprich2002}
		\item $^{2}$ \cite{Einasto2012} \hfill $^{b}$ \cite{Popesso2007a}
		\item $^{3}$ \cite{Pearson2014} \hfill $^{c}$ \cite{Boehringer2000}
        \item $^{4}$ \cite{Akamatsu2017} 

	\end{tablenotes}
\end{table*}

\begin{table*}
	\centering
	\caption{The mass complete non-merging cluster subset sample is presented here, noting the null hypothesis is accepted where $P(\Delta)\geq0.01$. All values and uncertainties are obtained and determined as detailed in Table \ref{tab:sub}.}
	\label{tab:nsub}
	\begin{tabular*}{\textwidth}{l@{\extracolsep{\fill}}cccccccr} 
		\hline
		Name & RA  & Dec & $L_{x}$  & $N_{r_{200}}$ & $\overline{cz}_{glob}$ & $\sigma_{r_{200}}$ & $\sigma_{ref}$ & $P(\Delta)$  
		\\ & (J2000) & (J2000) & ($\times 10^{44}$ erg s$^{-1}$) & & (km s$^{-1}$) & (km s$^{-1}$) & (km s$^{-1}$) &
		 \\ \hline
		 Abell 85 & 00 41 37.81 & -09 20 33 & 9.41$^{a}$ & 70 & 16709$\pm$71 & 719$^{+45}_{-55}$ & 979$^{5}$ & 0.853 
         
		\\ Abell 119 & 00 56 21.37 & -01 15 46 & 3.30$^{a}$ & 59 & 13279$\pm$74 & 752$^{+47}_{-59}$ & 619$^{6}$ & 0.579 
        
        \\ Abell 1650 & 12 58 46.20 & -01 45 11 & 6.99$^{a}$ & 50 & 25087$\pm$98 & 671$^{+58}_{-78}$ &  498$^{2}$ & 0.636
        
		\\ Abell 1656 & 12 59 48.73 & +27 58 50 & 7.77$^{a}$ & 145 & 6995$\pm$39 & 798$^{+27}_{-29}$ & 973$^{7}$ & 0.087 
        
		\\ Abell 1795 & 13 49 00.52 & +26 35 06 & 10.26$^{a}$ & 70 & 18754$\pm$87 & 794$^{+56}_{-69}$ & 662$^{2}$ & 0.265 
        
        \\ Abell 2029 & 15 10 58.70 & +05 45 42 & 17.44$^{a}$ & 127 & 23382$\pm 103$ & 932$^{+63}_{-79}$ & 973$^{7}$ & 0.415 
        
		\\ Abell 2061 & 15 21 15.31 & +30 39 16 & 4.85$^{d}$ & 90 & 23311$\pm$69 & 719$^{+43}_{-53}$ & 898$^{3}$ & 0.183 
        
        \\ Abell 2065 & 15 22 42.60 & +27 43 21 & 5.55$^{a}$ & 93 & 21565$\pm$92 & 882$^{+57}_{-72}$ & 1286$^{3}$ & 0.211  
        
		\\ Abell 2199 & 16 28 38.50 & +39 33 60 & 4.09$^{a}$ & 67 & 9161$\pm$55 & 737$^{+36}_{-42}$ & 722$^{6}$ & 0.586 
        
		\\ ZWCL1215 & 12 17 41.44 & +03 39 32 & 5.17$^{a}$ & 97 & 23199$\pm$98 & 671$^{+58}_{-78}$ & 889$^{9}$ & 0.873 
		\\
		\hline
	\end{tabular*}	
	\begin{tablenotes}
		\item $^{2}$ \cite{Einasto2012} \hfill $^{a}$ \cite{Reiprich2002} 
        \item $^{3}$ \cite{Pearson2014} \hfill $^{d}$ \cite{Marini2004} 
        \item $^{5}$ \cite{Agulli2016}
        \item $^{6}$ \cite{Rines2003}
        \item $^{7}$ \cite{Sohn2017}
        \item $^{8}$ \cite{Munari2014}
        \item $^{9}$ \cite{Zhang2011}
	\end{tablenotes}
\end{table*}


\section{Velocity Dispersion Profiles}
\label{sec:vdp}

The kinematics of each cluster within the sample are analysed from their respective velocity dispersion profiles, denoted as $\sigma_{P}(R)$. 
These VDPs can depict, with reasonable clarity, how dynamically complex or simple a cluster is. 
In this work we derive the VDPs computationally from the method prescribed by \cite{Bergond2006} for globular clusters. 
This has since been adapted to galaxy groups and clusters by \cite{Hou2009,Hou2012}. 
The VDPs are produced from bins of the radial velocities through a Gaussian window function that is weighted exponentially as a function of radius across all radii. 
However, in line with Harris (private communication), we note here the presence of a typographical error in the original notation of this function by \cite{Bergond2006}, in which the exponential component should be denoted to be negative rather than positive.
This error appears to have been perpetuated into further works cited here (e.g. \citealt{Hou2009}; \citealt{Hou2012}; \citealt{Pimbblet2014}).
We therefore present the corrected version of this window function in equation \ref{eq:weight}, which can be seen in the body of the work by \cite{Woodley2007} under equation 3.
The correct window function is written as

\begin{equation}
	\omega_{i}=\frac{1}{\sigma_{R}}\exp{-\Bigg[\frac{(R-R_{i})^{2}}{2\sigma_{R}^{2}}\Bigg]},
	\label{eq:weight}
\end{equation}

\noindent
where the kernel width $\sigma_{R}$ determines the size of a window that the radial velocities are binned against with the square-difference in radius $(R-R_{i})^{2}$. 
The window is chosen to be 0.2$R_{max}$ in units of $r_{200}$. 
This is to avoid the window being too large, thereby smoothing out features in the profile, or too small where spurious shapes in the profile could be produced by outliers. 
The window function $\omega_{i}$ is then applied to the projected VDP, which is written as

\begin{equation}
	\sigma_{P}(R)=\sqrt{\frac{\sum_{i}\omega_{i}(R)(x_{i}-\bar{x})^{2}}{\sum_{i}\omega_{i}(R)}},
	\label{eq:vdp}
\end{equation}

\noindent
where $x_{i}$ represents the radial velocity of each galaxy inputted taken as a difference from $\bar{x}$, which represents the mean recession velocity of the cluster. 

\begin{figure*}
	\centering
	\includegraphics[width=0.85\paperwidth]{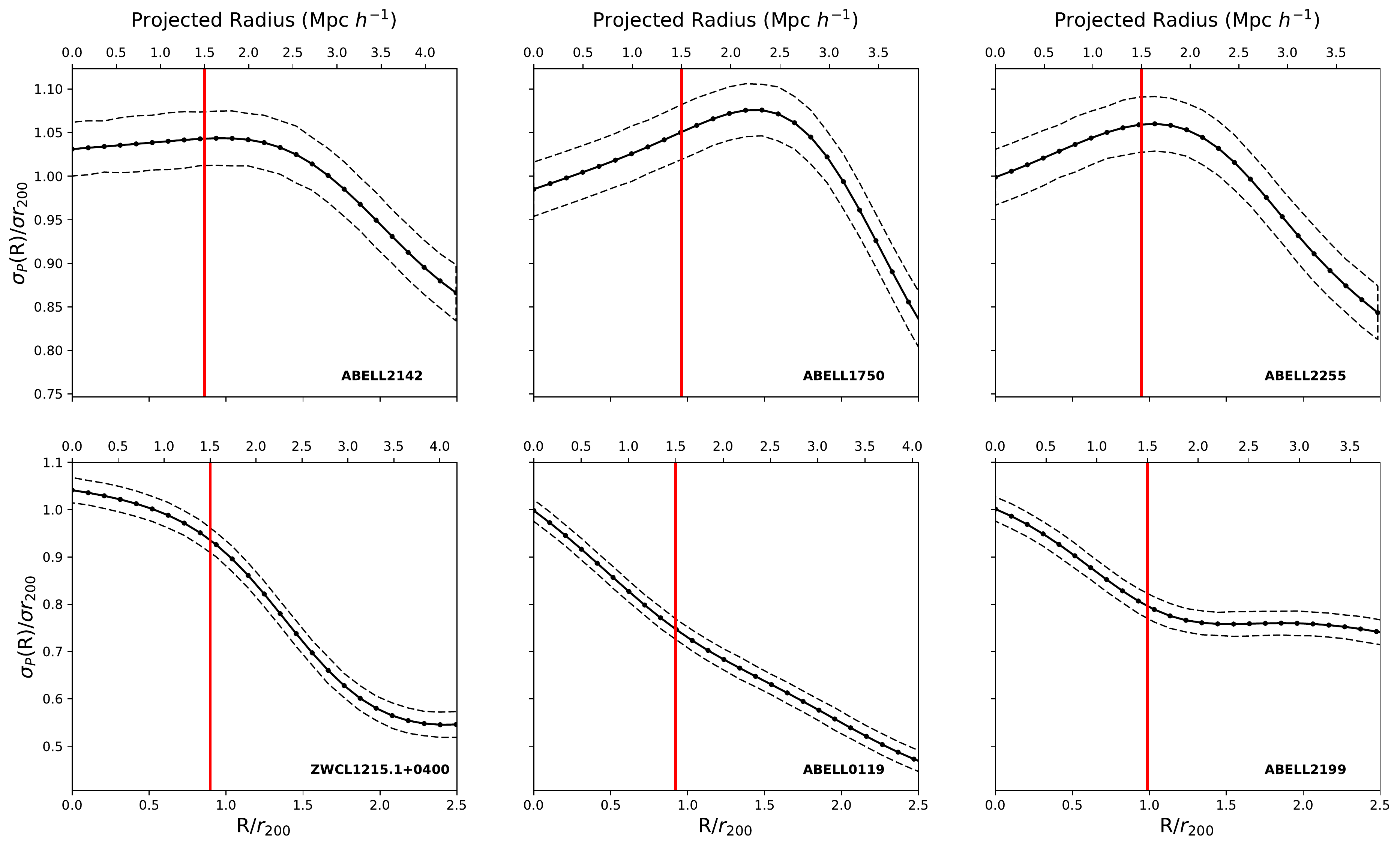}
	\caption{Example VDPs, consistent with those in Figure \ref{fig:caus}, from the merging (top row) and non-merging (bottom row) sub-samples plotted as a function of the projected virial radius $r_{200}$ and normalised to their respective central values. The red vertical line indicates $1.5$ Mpc $h^{-1}$ from the cluster centre where the global values and $\Delta$ test for sub-structuring are calculated. The dashed lines represent the 1$\sigma$ uncertainty of 1000 monte carlo resamples. Note the rising profiles within $1.5$ Mpc $h^{-1}$ in the merging clusters compared to the decreasing-to-flat profiles for the non-merging clusters within $1.5$ Mpc $h^{-1}$.}
	\label{fig:vdps}
\end{figure*}

The in-putted cluster data ideally should not have fewer than 20 galaxy members, this is to ensure the resultant projected VDPs are not specious \citep{Hou2009}. 
This can potentially pose problems for wanting to observe the dynamics of a cluster based on varying galactic parameters due to the inadvertent biasing to smaller bin sizes. 
Applying the cluster richness criterion of 50 galaxy members at $\leq r_{200}$ provides an adequate safeguard against this problem while determining cluster membership.
An example of the full non-split VDPs from each sub-sample are presented in Figure \ref{fig:vdps}.
From this we can see the bins that reside within $1.5$ Mpc $h^{-1}$ marry closely with the results from the $\Delta$ test for substructure, however, this is not found to be consistent across the entire the sample of determined merging and non-merging systems.
A problem which was noted by \cite{Pimbblet2014}, and could reflect the homogenisation of certain clusters where the weighting of the Gaussian moving window function causes a rise in response to more significant groupings of galaxies at larger radii.

\begin{figure*}
	\centering
	\includegraphics[width=0.85\paperwidth]{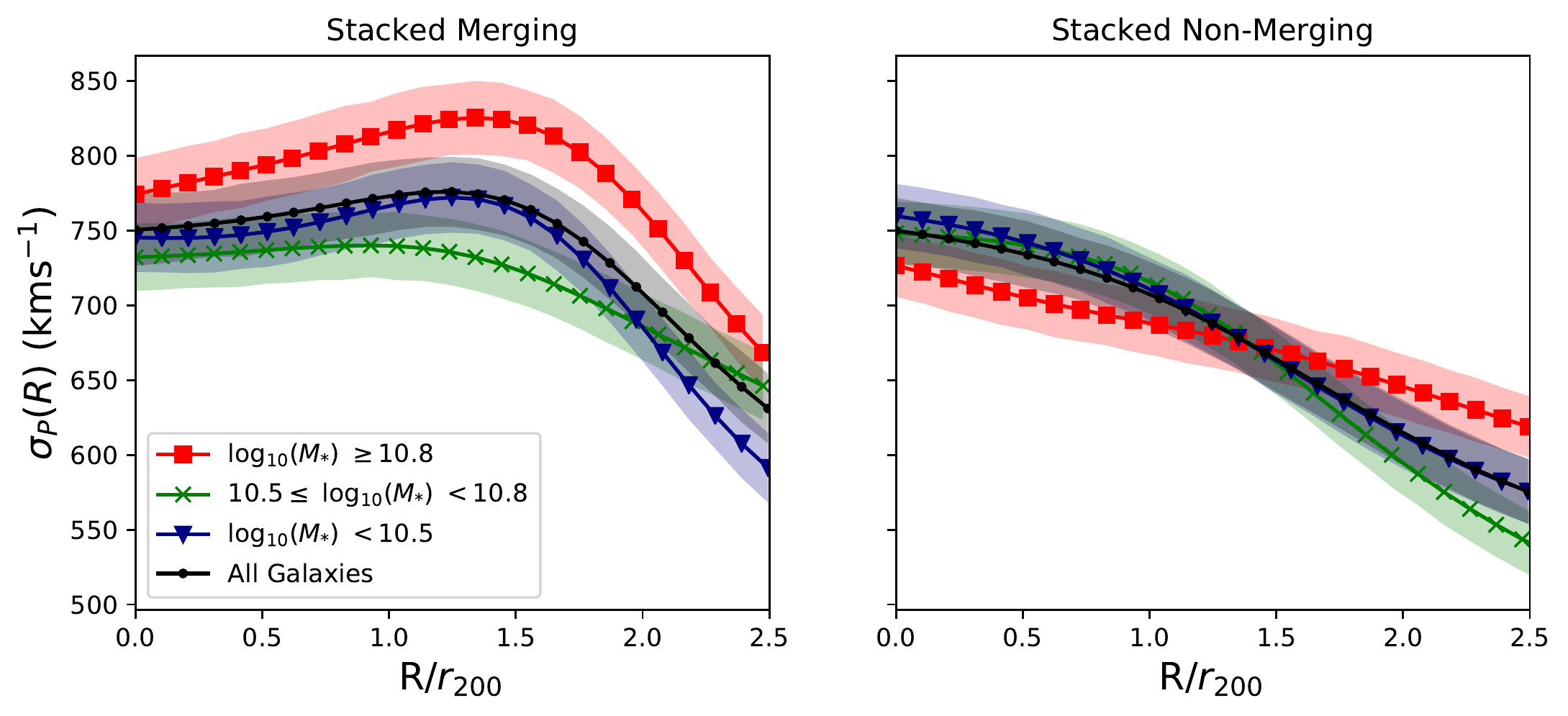}
	\caption{Co-added VDPs split by stellar mass for each cluster. Each profile represents a split by different intervals of log$_{10}$($M_{\ast}$) as a function of radius (R/$r_{200}$), with the black profile representing all available galaxies within the sample. Shaded regions represent the 1$\sigma$ uncertainty of 1000 monte carlo resamples.}
	\label{fig:stellar}
\end{figure*}

In order to address the aims of this work we compare the cluster environments between merging and non-merging systems with the kinematics of their member cluster galaxies through varying limits of different intrinsic cluster galaxy parameters. 
We therefore compute a series of VDPs with equations \ref{eq:weight} and \ref{eq:vdp} outlined in section \ref{sec:vdp} using the following methodology: 
Cluster galaxies are split between specific limits of varying galaxy properties of mass, morphology and colour. 
These splits are passed through each cluster from both samples, with each cluster galaxy co-added to a normalised fixed grid of line of sight velocity $\Delta V$, and projected radius $R$ between $0-2.5$ $r_{200}$. 
Resulting in a stack for each of the merging and non-merging samples. 
Stacking for each sub-sample allows for a general picture of each environment to be built, to ascertain how the kinematics of differing sub-populations of galaxies within each environment are affected.

\subsection{Mass}
\label{sec:splitmass}
Analysis of varying stellar mass limits allows for basic inference of how differing galaxy populations may vary depending on its environment at incremental radii from its centre. 
Fixed limits are chosen for 3 profiles of different masses: log$_{10}(M_{\ast})\geq 10.8$, $10.5 \leq$ log$_{10}(M_{\ast}) < 10.8$ and log$_{10}(M_{\ast}) < 10.5$. 
These limits are selected arbitrarily in order to maintain parity between the bin sizes of each range. 

Figure \ref{fig:stellar} shows the resultant stacks of the merging, and non-merging, clusters split via different stellar masses present in the DR8 data.
In the merging stack, there is a prominent illustration of a dynamic environment, especially between the log$_{10}$($M_{\ast}$) $< 10.5$ and log$_{10}$($M_{\ast}$)$\geq 10.8$ profiles.
The log$_{10}$($M_{\ast}$) $\geq 10.8$ mass profile shows a steadily increasing profile to having the highest dispersion of velocities at $\sim1.5 r_{200}$, in tandem with the log$_{10}$($M_{\ast}$)$< 10.5$ profile.
The former commonly denoted as members of an accreted older population of galaxies, with the latter commonly associated with an accreted younger population.
The log$_{10}$($M_{\ast}$)$\geq 10.8$ profile represents an increasing intensity of interacting, or merging, galaxies at $\lesssim 1.5$ $r_{200}$. 
The same can be determined with the log$_{10}$($M_{\ast}$) $< 10.5$ mass profile, which demonstrates a level of merging activity that is in tandem with the \textquoteleft All Galaxies' profile peaking at $\sim 1.5$ $r_{200}$.
These are clearly the two prominent sub-populations that drive the dynamic nature of the merging stack.
The \textquoteleft All Galaxies' profile shows a parity with the log$_{10}$($M_{\ast}$)$< 10.5$ profile throughout, suggesting the lower mass galaxies dominate the kinematics of the stack.
At $\sim 1.5$ $r_{200}$ it appears there is a high level of mixed substructuring between the log$_{10}$($M_{\ast}$) $< 10.5$ and log$_{10}$($M_{\ast}$)$\geq 10.8$ populations.
Where the \textquoteleft All Galaxies' profile seems to indicate it is primarily composed of the two aforementioned sub-populations at larger radii.
This is indicative of the occurrence of pre-processing by accretion of galaxies onto groups prior to cluster accretion.
The intermediate profile of 10.5 $\leq$ log$_{10}(M_{\ast}) < 10.8$ is the flattest, therefore, least dynamic of the populations within the stack.
The non-merging sample is comparatively dynamically relaxed with smaller dispersions and declining profiles that are not too dissimilar to the trend shown by \cite{Girardi1996}.
The log$_{10}$($M_{\ast}$) $< 10.5$ shows the closest parity with the \textquoteleft All Galaxies' profile stack, again, suggesting low mass galaxies dominate the kinematics.
This profile possess the highest dispersion of velocities within $r_{200}$, indicative of a young infalling population of galaxies.
Whereas the log$_{10}$($M_{\ast}$)$\geq 10.8$ profile has the lowest dispersion within $r_{200}$.
This could be an indicator of an old population of galaxies slowly sloshing with the recently collapsed members onto cluster potentials.
The 10.5 $\leq$ log$_{10}(M_{\ast}) < 10.8$ profile blends in with the \textquoteleft All Galaxies' and log$_{10}$($M_{\ast}$) $< 10.5$ profiles, suggesting there is little diversity between these populations of galaxies.

\subsection{Colour}
\label{sec:splitcl}

\begin{figure}
	\includegraphics[width=\columnwidth]{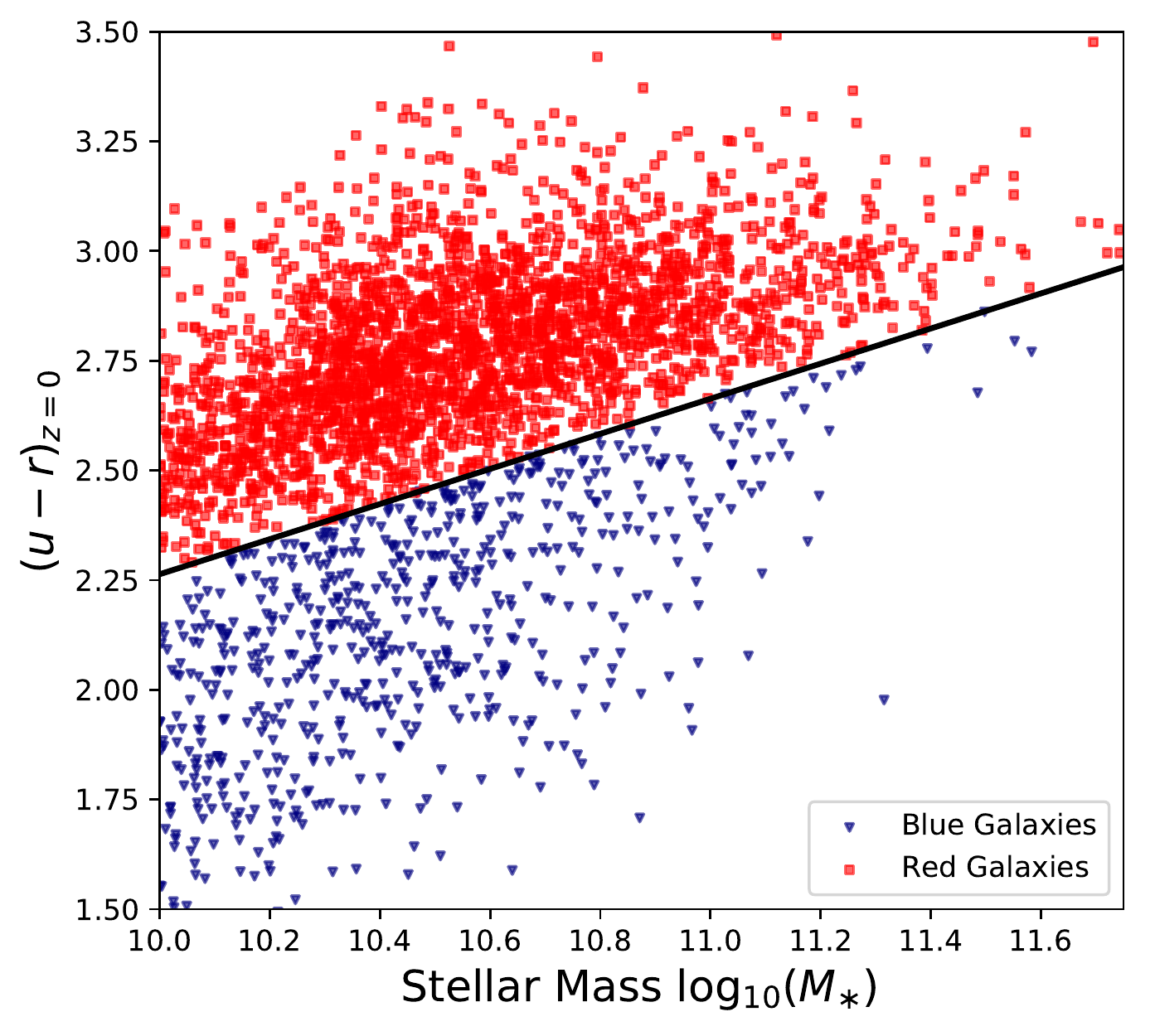}
    \caption{$(u-r)_{z=0}$ plotted as a function of log$_{10}$($M_{\ast}$). The black line resembles the linear fit of the centre of the bimodal distribution at quartile increments of log$_{10}$($M_{\ast}$); red galaxies are above the fitted line denoted as red squares; blue galaxies are below the fitted line denoted as blue triangles.}
    \label{fig:cldist}
\end{figure}

The cluster galaxies of each sample are passed through a colour limit gradient as a function of stellar mass. 
This is determined through the residuals of the bimodal distributions of colour in quartile increments of stellar mass (see \citealt{Jin2014}). This results in the following linear relation

\begin{equation}
	(u-r)_{z=0} = 0.40[\log_{10}(M_{\ast})]-1.74,
	\label{eq:clrel}
\end{equation}

\noindent
which as a consequence allows for an adequate boundary between red and blue galaxies that accounts for the biasing of galaxy colour distributions between low and high stellar masses. 
The resultant boundary line and the galaxy distributions can be seen in Figure \ref{fig:cldist}.

It should be noted that not all galaxies possess the used \texttt{\textquoteleft modelMag'} DR8 photometry, therefore, some clusters experience a slightly reduced bin size compared to the principle MPA-JHU derived parameters. The galaxy colours are k-corrected to $z=0$ prior to computing the VDPs with the imposed variable limit (see \citealt{Chilingarian2010}; \citealt{Chilingarian2012}). 

\begin{figure*}
	\centering
	\includegraphics[width=0.85\paperwidth]{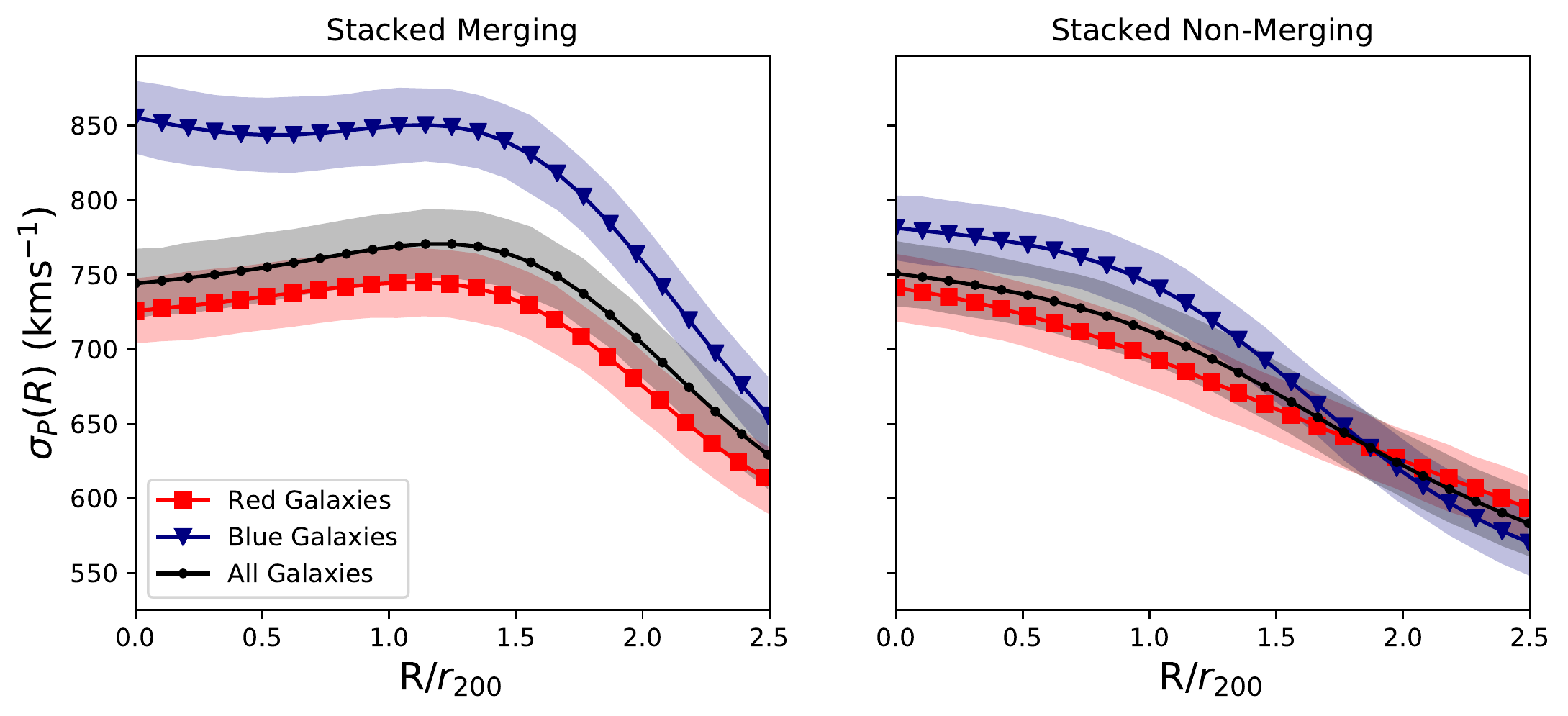}
	\caption{Stacked galaxy cluster VDPs split by their colour with the same axes as Figure \ref{fig:stellar}. Where the blue triangle and red square profiles represent the blue cloud and red sequence respectively, with the black dot profiles representing all cluster galaxies available with colour data. Shaded regions represent the 1$\sigma$ uncertainty of 1000 monte carlo resamples.}
	\label{fig:cl}
\end{figure*}

Figure \ref{fig:cl} depicts the merging sample to have a consistently high dispersion profile for the blue cloud stack at $\leq1.5$ $r_{200}$, where the red sequence presents a shallow rising VDP with radius. 
However, there is a rising kinematic parity of the red sequence profile with the \textquoteleft All Galaxies' profile throughout $\leq1.5$ $r_{200}$.
This behaviour could be an indicator the population of blue, presumably star-forming, galaxies are kinematically active due to pre-processing of galaxies within the merging substructure with gradual infall onto the potential; \cite{Haines2015} highlights the need of pre-processing galaxies into groups to account for the level of quenching of star formation observed in cluster galaxies at large clustocentric radii.
The rising profile of the red population with radius potentially demonstrates another environment of interacting galaxies, the profile leads to a rising VDP, indicating groupings of red galaxies at larger radii.
These results evince a mixed population of merging blue galaxies alongside already pre-processed red galaxies in sub-groupings at larger radii from the cluster centre.
The non-merging sample shows less dynamical variation, where all of the profiles present a shallow-to-flat variance with $R/r_{200}$.
The shallow rising of the blue galaxy profile at $\sim1.5$ $r_{200}$ could be an indicator of an infalling population of blue galaxies, that have not tidally interacted with other cluster members to the same degree as the merging counterpart.
Comparatively, the red population profile presents gradual decrease from faster velocity dispersions at $\leq$ $r_{200}$.  
There is the conspicuous observation of the merging red VDP in Figure \ref{fig:stellar} representing high mass galaxies in that it does not marry with what we would anticipate in comparison the merging red VDP in Figure \ref{fig:cl} representing red galaxies.
However, the mass limits in section \ref{sec:splitmass} are independent of colour, therefore, there is a mix of red and blue galaxies in the high mass sample of galaxies. 
This is combined with a discrepancy in the sample sizes between a bi-modal colour split and that of stellar mass which can be seen in Figure \ref{fig:cldist}, which is indicative that the red low-to-intermediate mass galaxies contribute to lower velocity dispersions.
This behaviour does match with what \cite{Girardi1996} believe to indicate a neighbouring system or grouping of galaxies at larger radii.
The direct comparison between the merging and non-merging samples in Figure \ref{fig:cl} demonstrates a more diverse variation of colour in dynamically relaxed clusters when compared to those that are dynamically complex, which has been discussed with recent observations made by \cite{Mulroy2017}. 

\subsection{Morphology}
\label{sec:splitmorph}

The morphological classification of galaxies in clusters can be used as a proxy on how the local environment can lead to an alteration of their structure and shape. 
Therefore, utilising the debiased morphological classification data of GZ2, this is married with the data of both merging and non-merging samples split by the same colour limits noted in \ref{sec:splitcl}. 
The samples are separated between umbrella spiral and elliptical morphologies, which is determined using the string classifier of \texttt{\textquoteleft gz2\_class'} by whether or not a galaxy possessed any number of spiral arms in its structure. 
It should be noted, however, that the relatively small number of galaxies classified within GZ2 ($\sim300,000$) means the average cluster membership can drop significantly. 
As a result the two clusters, Abell 0426 and Abell 0085, are not added to the stack for not meeting the richness criteria highlighted in section \ref{sec:data}.
This drop in membership could lead to the average profiles being spurious due to the lack of a more complete data set. For each morphology in each environment, the cluster galaxies are then split into the same colours via the same linear relation as noted in section \ref{sec:splitcl}.

\begin{figure*}
	\centering
	\includegraphics[width=0.85\paperwidth]{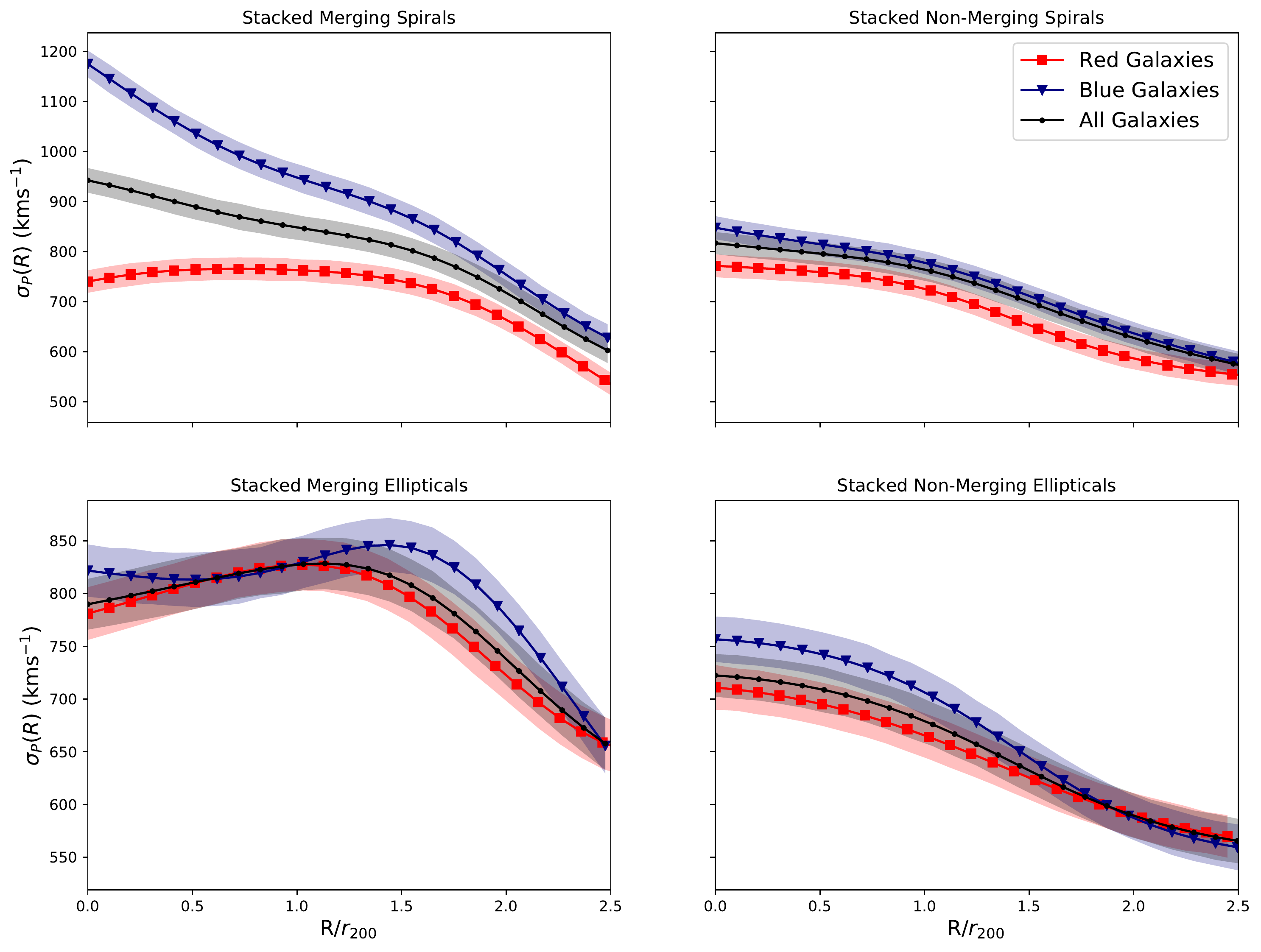}
	\caption{Co-added VDPs of spirals and ellipticals for each of the individual environments, which are then split by their bi-modal colours as per Figure \ref{fig:cl}. Shaded regions represent the 1$\sigma$ uncertainty of 1000 monte carlo resamples.}
	\label{fig:morphcl}
\end{figure*}

Figure \ref{fig:morphcl} presents the resultant morphology-colour split. 
The merging spiral stack shows a declining blue population profile, that then converges with the \textquoteleft All Galaxies' profile. 
This coincides with a near flat profile of the spiral red population that starts to decline at $\sim r_{200}$. 
It is clear the merging spiral blue and red populations equally contribute to the total dispersion of merging spirals of the \textquoteleft All Galaxies VDP. 
However, there is a discrepancy from the conspicuously high dispersion of blue spirial galaxies within $r_{200}$, suggesting there is an infalling, or recently accreted, high velocity population of blue spirals. 
The blue population profile of the merging ellipticals is fairly dynamic, leading to bulk rise at $\sim1.5$ $r_{200}$. 
This is indicative of a strongly interacting sub-population of cluster galaxies, potentially as the result of tidal-tidal interactions through substructuring. 
The red elliptical profile, which shows a bulk rise at $\sim1.2$ $r_{200}$, reaching close parity with the \textquoteleft All Galaxies' profile, indicating the red ellipticals be the main contributor to the \textquoteleft All Galaxies' profile. 
The red ellipticals, like the blue ellipticals, present an interacting sub-population within a merging environment, potentially these could be older pre-processed galaxies that were harassed into substructures at a subtly earlier epoch.
The merging elliptical VDPs consist of mixed blue and red elliptical galaxies that have gone through pre-processing interactions beyond $r_{200}$
Both colour sub-populations in the merging elliptical stack are consistent with the blue and red merging sub-populations in Figure \ref{fig:cl}, insinuating that ellipticals are the dominant contributors to a merging cluster environment.

In contrast with the non-merging sample, the spiral galaxies of both colour sub-populations steadily decline with radius.
The non-merging ellipticals present a similar uniform of profiles that steadily decline with radius, aside from the slight increase in the dispersion of blue ellipticals at $\lesssim r_{200}$ suggesting they are recent members to collapse onto the cluster potential. 
The general slow decline observed with these non-merging profiles indicates a comparatively mixed ambient system of cluster galaxies.
The merging VDPs are overtly dynamic, especially with the high dispersions in blue spiral cluster galaxies, or the variable profile shapes in the ellipticals, when compared to their non-merging counterparts.
This is a clear indication of the differences in dynamical ages of the two environments; active feeding of a cluster potential through substructuring and infall compared to one which has reached a relaxed dynamical equilibrium.

\section{Discussion}
\label{sec:disc}

The work presented here shows that across all intrinsic galactic parameter splits, the merging samples possess some form of rising profile.
\cite{Hou2009} argues that such a rise indicates an interacting, or merging, system based on a correlation between a sample of non-Gaussian galaxy groups, coinciding with previous work by \cite{Menci1996}. However, these earlier works did not explicitly delineate which class(es) of galaxy are driving this.

\subsection{Interpreting the VDPs}
\label{sec:ivdp}

When analysing the \textquoteleft All Galaxies' profiles for each split of the merging stacks, it can be deduced that these results seemingly back the argument made by \cite{Menci1996} and \cite{Hou2009}. 
With the non-merging samples generally showing a flat-to-declining series of profile.
These results could corroborate recent work by \cite{Mulroy2017} that finds different cluster evolutionary histories must have played a part to explain the prominent colour variation observed in non-merging systems compared to that of merging systems. 
\cite{Deshev2017} is consistent with this, observing a significant decrease in the fraction of star-forming galaxies in the core of the merging Abell 520 system compared against their non-merging sample, with evidence of a smaller group of galaxies, possessing a higher fraction of star forming galaxies, feeding the merger. 
One explanation for this observation suggests a non-merging galaxy cluster is formed on long timescales by their haloes inducing the infall, and accretion, through harassment of galaxies from the surrounding field population that leads to the gradual variation from red to blue colours with increasing radius from the centre seen in \cite{Mulroy2017}.
Whereas the merging systems are formed primarily from the accretion of pre-processed galaxy groups, meaning the galaxies have undergone heavy interactions leading to evolutionary changes, and are virialised to their local groupings. 

We find the red populations of the merging stacks are the main contributors to the rising profiles, which illustrates a common and significant amount of interactions occurring at $\sim1.5$ $r_{200}$ radii. 
Although, consideration should be taken into account that red galaxies could overshadow the total colour distribution of the cluster galaxy sample by numbers alone due to the Malmquist bias \citep{Malmquist1925}, along with the making the sample complete, thereby impeding a true indication on how these two sub-populations behave kinematically.
In comparison to the non-merging profiles that clearly illustrate a more relaxed environment with a possible suggestion of infalling blue galaxies, this married with the merging profiles showing the dominant driver of the rising profile shape to be a mix of red and blue elliptical sub-populations. 
The diverse dynamics between merging and non-merging systems provide further affirmation to the idea of a galaxy infall and accretion bi-modality between merging and non-merging systems.

Considering the epochs of differing events that occur in a typical cluster (e.g. infall, accretion, splashback), we can use the timescales between them to try and infer the current physical processes occurring and how they relate to their kinematics. 
\cite{Haines2015} simulate the accretion paths of multiple galaxies onto a massive cluster from various epochs and classify the infall regions to start $\lesssim 10 h^{-1}Mpc$, or $\lesssim 5 r_{200}$. 
It is calculated that the timescales from infall to accretion to be $\sim 4$Gyr, a galaxy then becomes accreted once it reaches $r_{200}$ and passes its first pericentre on timescales of $0.5-0.8$Gyr, followed by a significantly slower of $2-3$Gyr for the galaxy to reach its first apocentre (splashback radius). 
Collectively, the VDPs demonstrate a period of infall in the merging stacks at $\leq r_{200}$, alongside a culmination of interactions occurring as a result of the domination of pre-processed groups. 
This is corroborated with the merging colour and morphology VDPs, where mixed blue and red populations of galaxies assumed to be undergoing pre-processing are infalling to be accreted onto the cluster, reaffirming the suggestion by \cite{Haines2015} that pre-processing is required to explain star formation being quenched at larger radii from the cluster centre. 
Furthermore, the VDPs representing spiral morphology could be indicating the galaxies at $\gtrsim1.25$ $r_{200}$ are the start of a $\sim4$Gyr long journey onto the cluster potential, leading to their accretion and possible splashback, thus accounting for the larger surface density of spirals at smaller clustocentric radii (see \citealt{Wetzel2012}; \citealt{Haines2015}; \citealt{Cava2017}). 

In any case, there are increasingly more observations and simulations that appear to occasionally contradict, where many authors suggests a need for pre-processing (\citealt{Haines2015}; \citealt{Roberts2017}; \citealt{Carvalho2017}). 
\cite{Mulroy2017} argues for a bi-modality on infall and accretion histories with similar accretion rates, one with pre-processing and one without, in order to explain the variations in colour found in non-merging systems. Further simulations could possibly help to build on this picture for these bi-modal, kinematic outcomes.

\subsection{Phase-Space Caustics}
\label{sec:psc}

In Section \ref{sec:data} we calculate velocity dispersions through a biweight method \citep{Beers1990} and the phase-space surface caustics to determine cluster membership (\citealt{Diaferio1997}; \citealt{Diaferio1999}).
The phase-space caustics produced from the chosen methodology follow a trumpet-shape pattern as we move away from the cluster centre, which is a result from galaxies infalling onto the cluster when the potential inundates the Hubble flow \citep{Regos1989}.
\cite{Diaferio1997} and \cite{Diaferio1999} both demonstrate the amplitudes of these surface caustics to be a product of random non-radial motions from substructuring, indicating a diverging caustic to be illustrative of a cluster with increasing interactions. 
Therefore, these caustics represent an escape velocity of the cluster potential.
The key benefit, aside from powerfully indicating cluster boundaries, is that these caustics can be produced on redshift data alone.
Unlike the rest of the literature, we allow the surface caustics to stretch to a $\Delta$V velocity limit of $\pm 1500$ kms$^{-1}$.
This is to allow infallers to be added into the sample of cluster galaxies for each cluster, although, we wish to note that this method involves the risk of adding interloping larger scale structures to the sample.
Many of the clusters compiled within this sample have been well studied, with calculated surface caustics and velocity dispersions.
Reference values for the latter are presented in both Tables \ref{tab:sub} and \ref{tab:nsub}.
The calculated $\sigma_{r_{200}}$ velocity dispersions are fairly consistent with the reference literature, however, there will be differences dependent on which method was used to estimate the velocity dispersions, at what radial point and how many galaxies are available for the membership of the cluster at $\leq r_{200}$ in this work.
What follows is a comparison of our phase-space surface caustic analysis with that of the literature.

\subsubsection{Abell 85}

Abell 85 is a well studied cluster, with multiple calculations of its dispersion of velocities, along with phase-space surface caustics presented within \cite{Rines2006}. 
The value of $\sigma_{r_{200}}$ from this work is $\sim200$ kms$^{-1}$ offset from the calculated literature values. 
The primary driver of this offset is due their cluster membership being significantly greater with 497 galaxies within 1.7 $r_{200}$ compared to 234 galaxies within 2.5 $r_{200}$ from the data used here.
The vast difference in galaxy membership can induce a slight alternate shape between the resultant surface caustics.
\cite{Agulli2016} do not publish the surface caustics on their phase-space diagrams, leaving the surface caustics of \cite{Rines2006}, which indicate a strong constraint in $\Delta V$-space at low radii.
Despite the lack of sharp, sudden changes in the surface caustic with increasing R, there are still similarities in the membership from the caustic presented here against that of \cite{Rines2006}.
This indicates there is consistency between the two independent calculations of the caustic surface that allows for a more liberal inclusion of galaxies into the membership.

\subsubsection{Abell 119}

Abell 119 possesses multiple surface caustics in the literature alongside calculations of their velocity dispersions (\citealt{Rines2003}; \citealt{Rines2006}).
There is, again, an offset of $\sim100$ kms$^{-1}$ in the calculation of $\sigma_{r_{200}}$, for which similar reasoning is applied from that of our discussion on Abell 85; the radial point at which the velocity dispersion is calculated can push the gaps between the literature further.
Additionally, the techniques used for calculating the velocity dispersion from this work varies from that of \cite{Rines2006}, where sigma clipping is used \citep{Zabludoff1990}, this will lead to an underestimating of the velocity dispersion when directly compared to a biweight estimator.
The phase-space caustics are the most consistent with the CAIRNS cluster study of \cite{Rines2003}, with very similar profiles.
These caustics only deviate where there are discrepancies in the number of galaxies within $\leq10$ Mpc $h^{-1}$.
The recalculated caustics presented in \cite{Rines2006} focus on constraining the cluster membership by limiting galaxies in $\Delta V$-space to $\leq 1000$ kms$^{-1}$, creating a surface caustic that is not as smooth, but is effective in the elimination of infallers and the encompassing large scale structure.

\subsubsection{Abell 426}

Abell 426, commonly known as the \textquoteleft Perseus cluster', does not presently possess any phase-space caustic analysis in the literature.
Although, the phase-space surface caustics determined here are relatively simple, and the population of galaxies accumulated does not extend beyond $\sim2$ Mpc$h^{-1}$, providing a smooth distribution with several groupings of member galaxies.
The limited and immediate break in the available data, due to the survey's limitations in observing the north galactic cap, lends to an artificial increase in the VDP at larger radii.
However, this affect should be reduced when stacked against the other clusters that extend beyond the projected radii of Abell 426.
The velocity dispersions of Abell 426 determined within this work are not consistent with those determined within the literature, showing an offset of $\sim500$ kms$^{-1}$ \citep{Struble1999}.
The lack of consistency is a result of the significant loss of galaxy members compared to the true scale and size of Abell 426, which contains close to $\sim1000$ galaxy members.

\subsubsection{Abell 1650}

Abell 1650 is an atypical cluster with a radio quiet cD cluster galaxy at its centre. 
The surface caustics presented in the literature follow \citep{Rines2006} a similar shape and profile to our surface caustics, with a slight difference to the radial cut used on the sample of galaxies and a wider velocity window to allow for the addition of galaxy infallers.
The velocity dispersions produced within this work are consistent with those of \cite{Einasto2012}, within a slight discrepancy of $\sim200$ kms$^{-1}$.
Although, the discrepancy in these values is expected due to differing methods used in calculating the dispersion.

\subsubsection{Abell 1656}

Abell 1656, commonly referred to as \textquoteleft Coma', is a well studied cluster with close to $\sim1000$ members. 
It has such a strong presence within the literature primarily due to its relatively close proximity (z $\sim0$), which results in a greater sacrifice of cluster galaxies when maintaining completeness.
However, this is offset by the extremely high number density of cluster galaxies.
The phase-space caustics of the coma cluster presented in this work are the most consistent with \cite{Sohn2017}, this is the result of a more relaxed $\Delta V$-space limit to accommodate the very large nature of the cluster.
This consistency is lost at $\sim$ 4 Mpc $h^{-1}$ due to a sudden drop in galaxies present within our MPA-JHU sample.
However, an assumption can be made based the consistency is valid due to the trend of the caustic profile following that of \cite{Sohn2017} closely.
The same consistency exists for the values of the velocity dispersion with very small offsets when compared to values from the literature (\citealt{Rines2003}; \citealt{Sohn2017}).

\subsubsection{Abell 1750}

Abell 1750 is a complex triple subcluster system in a pre-merger state, which is briefly discussed in \ref{sec:subs}.
The phase-space surface caustics presented here are the most consistent with produced by \cite{Rines2006}, with the exception of allowing infallers at $\sim$2 Mpc $h^{-1}$ to form the cluster membership.
The literary values of the velocity dispersion show a discrepancy of $\sim100$ kms$^{-1}$ from the values calculated in this work (\citealt{Rines2006}; \citealt{Einasto2012}).
What does remain consistent is the reasoning that alternative, less robust, methods were used to calculate a value for $\sigma$.
As well as this, there is a lack of clarity on the exact methodology used to calculate the dispersions of velocities within some of the literature where alternative limits could have been used within their calculations that are otherwise unstated.

\subsubsection{Abell 1795}

Abell 1795 is a cool core galaxy cluster with an unusually large cavity with no counterpart \citep{Walker2014}.
There is currently no phase-space surface caustic analysis within the literature that can be aided to check consistency.
However, from our own determined caustics we can see there is a roughly even distribution of member galaxies close to the centre of the cluster, as expected from a typical relaxed cluster.
Our calculated velocity dispersion is consistent with those values found in the literature (\citealt{Zhang2011}; \citealt{Einasto2012}).

\subsubsection{Abell 2029}

Abell 2029 is a massive cluster that possess a powerful cD galaxy at its centre, forming part of a supercluster with complex dynamical interactions within the ICM \citep{Walker2012}.
\cite{Sohn2017} has produced surface caustics of Abell 2029 that are inconsistent with our own.
There are gaps in the galaxy population size within the phase-space diagram due to the redshift limitations of the MPA-JHU DR8 data. 
These limitations make our data incomplete for this cluster, whereas \cite{Sohn2017} has used complementary sets of data, and therefore, does not possess the same restrictions as those found in this work.
However, the bulk of the galaxies present within the imposed limits of this work match those defined as members within the phase-space surface caustic diagrams of \cite{Sohn2017} that include infallers.
The calculated velocity dispersion is calculated in this work is consistent with other determined values within the literature despite the variances in galaxy membership. 

\subsubsection{Abell 2061}

Abell 2061 is a double subcluster system with complex dynamics that is in close proximity to Abell 2067, this is highlighted in more detail in \ref{sec:subs}. 
The comprehensive CIRS survey by \cite{Rines2006} presents consistent phase-space surface caustics when in consideration for the discrepancy in the range of velocities used.
The only discrepancy of note is the the presence of strong foreground substructuring at $\sim3.5$ Mpc $h^{-1}$ inducing the caustic profile to maintain a consistent velocity of $\sim1000$ kms$^{-1}$, which causes the VDP to slight increase beyond the $\sigma_{r_{200}}$ values.
The literary values for Abell 2061's velocity dispersion are consistent with our own where \cite{Pearson2014} presents an offset of $\sim100$ kms$^{-1}$, however, this is primarily due to the tighter distribution of galaxies, as well as differing methodologies for calculating the dispersion. 

\subsubsection{Abell 2065}

Abell 2065, at present, does not have any detailed phase-space analysis within the literature for direct comparison.
However, from our own analysis, Abell 2065 possesses what appears to be a strong bi-modal distribution, which can be attributed to a complex dynamical system of multiple substructures.
This would provide consistency, since Abell 2065 is stated in the literature to possess an unequal core merger, for which the full nature of this is detailed in \ref{sec:subs}.
We believe the relatively flat velocity offset at $\sim$ $-2000$ kms$^{-1}$ with increasing R to be the smaller of the two cores.
The state of initial merger makes it difficult for the surface caustics to discern where the cluster ends and begins.
However, the string of flat galaxies implies something akin to the Kaiser effect \citep{Kaiser1987}, where a flat radial separation against a non-flat separation in the plane of the sky leads to the inference of infallers.

\subsubsection{Abell 2142}

Abell 2142 is a notorious cluster for its smooth and symmetric X-ray emission, indicative of a post core-merger event, which occurred $\sim1$ billion years ago \citep{Markevitch2000}. 
The phase-space surface caustics of Abell 2142 vary within the literature, as well as in comparison to the work done here.
\cite{Munari2014} presents surface caustics within the confines of $\sim3$ Mpc $h^{-1}$ and appear to be constant with increasing $R$.
Again, with \cite{Rines2006} demonstrating a more dynamic and tighter caustic due to differing limits applied in both velocity-space and radial-space alongside data visualisation effects. 
As usual, the shapes of these caustics are determined by the numbers of galaxies present within the field and how closely, or sparsely, they are distributed as we increase $R$ from the cluster centre.
Again, the calculated velocity dispersions from \cite{Munari2014} are inconsistent with our own value, offset by $\sim300$ kms$^{-1}$.
This is due to the spread, number and density of the cluster membership determined in the work of \cite{Munari2014} being equally greater.

\subsubsection{Abell 2199}

Abell 2199 is a relatively local galaxy cluster and provides a good testing-bed for large scale structure formation thanks to its close proximity, this is akin to Abell 1656, another relatively local cluster.
The cluster is well studied, possessing several phase-space surface caustics in the literature.
The phase-space caustics in this work are the most consistent with \cite{Song2017} and \cite{Rines2003}, where the shape and profile closely matches despite a lower membership.
The velocity dispersions calculated here are consistent with those found within the literature \citep{Rines2003}.

\subsubsection{Abell 2255}

Abell 2255 is a merging galaxy cluster with a complex X-ray distribution, which has yielded a variety of studies to better understand the mechanisms of diffuse radio emission \cite{Akamatsu2017}.
The total membership of Abell 2255 in this work is considerably less than that of other literature.
However, the surface caustics of this work are still reasonably consistent with the caustics determined by \cite{Rines2006}, if lacking in definition.
The velocity dispersion profiles determined here are consistent with those in the literature, despite offsets of $\sim200$ kms$^{-1}$, the drivers are variations in galaxy membership (\citealt{Zhang2011}; \citealt{Akamatsu2017}). 

\subsubsection{ZWCL1215}

The phase-space caustics of galaxy cluster ZWCL1215 determined in this body of work is consistent with those that are produced by \cite{Rines2006}, with only slight variations in the definition of the shape of the surface caustics.
The calculated velocity dispersions are also consistent with those determined by \cite{Zhang2011}, with an offset of $\sim200$ kms$^{-1}$, as a result of the reduced membership of galaxies presented within this work.

\subsection{Interloping Structures}
\label{sec:subs}

The clusters that form our sample are not purely isolated potentials, therefore we should take into consideration potential interloping structures as a result of a cluster being a member of supercluster.
As an example, during the data accumulation stage of section \ref{sec:data}, the clusters are cross matched against the \cite{Einasto2001} catalogue of superclusters to determine any significant contamination between clusters.
Abell 2244 and Abell 2249 are eliminated from the samples due to their strong interloping/overlap in RA-DEC space and z-space within the regions being investigated within this work.
Although, their removal from the samples has not altered to shape of the final stacked VDPs to any significant degree.

There are also other clusters within the sample that possess unusual substructures.
The phase-space diagram of Abell 2065 in Figure \ref{fig:caus} clearly presents two seemingly independent structures. 
However, Abell 2065 has been documented in the literature to be at the late stage of an ongoing merger \citep{Markevitch1999}. 
Further X-ray observations with XMM-Newton indicate more specifically the presence of an ongoing compact merger between two subclusters within Abell 2065, where the two cores are at an epoch of initial interaction \citep{Belsole2005}. 
Higher resolution X-ray observations from Chandra show a surviving cool core from the initial merger, with an upper limit merger velocity of $\lesssim$1900 kms$^{-1}$, adding to the argument that Abell 2065 is an unequal core merger (see \citealt{Chatzikos2006}).
This provides an explanation to the slightly off-centre line-of-sight mean velocity distribution of galaxies, with a second, smaller core averaging out to $\sim$ $-1500$ kms$^{-1}$ found in the phase-space diagram of Abell 2065, and naturally will affect the shape of the VDP at larger radii.
Abell 1750 is a triple subcluster system with the north subcluster separated from the central subcluster by a velocity offset of -900 kms$^{-1}$ and are all currently in a stage of pre-merger to the point where the plasma between the substructures is significantly perturbed (\citealt{Molnar2013}; \citealt{Bulbul2016}).
In contrast Abell 2061, which resides within the gravitationally bound Corona Borealis supercluster with Abell 2065 (see \citealt{Pearson2014}), possesses two optical substructures that will affect the VDP similarly to Abell 2061 \citep{Weeren2011}.
It should be noted that Abell 2061 potentially forms a bound system with the smaller cluster/group Abell 2067 (\citealt{Marini2004}; \citealt{Rines2006}), with line-of-sight velocity separation of $\sim725$ kms$^{-1}$ \citep{Abdullah2011}.
Observations hint at a likely filament connecting the two systems \citep{Farnsworth2013} aiding to the suggestion of cluster-cluster interloping.
There is $\sim30'$ of sky separation and with the prescribed cosmology in section \ref{sec:intro} this provides a rough projected distance of $\sim$2.7 Mpc $h^{-1}$ from the centre of Abell 2061.
Yet, this confirms to the cluster-cluster overlapping suggestion with the criteria used to develop cluster membership.
Therefore, it is very likely the membership of Abell 2061 is contaminated with the infalling Abell 2067 cluster's member galaxies as we approach $2.5$ $R/r_{200}$.

\subsection{The Delta Test}
\label{sec:DT}

The process of determining whether or not a cluster is merging involved the use of the $\Delta$ test for substructure, devised by \cite{Dressler1988}. 
Whereby the presence of any substructure to a $\geq99\%$ significance is recorded as a merging cluster environment.  
The $\Delta$ test, while a powerful and sensitive tool, is limited in its power to test for substructure since it only concerns itself with the sum of the deviations of a local velocity dispersion and mean recession velocity with global cluster values. 
This could lead to a greater probability of false positives for sub-structuring, along with omissions of those clusters that genuinely possess it. 
The problem becomes more apparent if an appropriate radial cut-off is not applied when calculating $\Delta$, otherwise the test will classify nearly every cluster to contain substructure.
This is a consequence of the varying numbers of cluster galaxies that are added into the calculation of $\Delta$; greater numbers of cluster galaxies help decrease the value of $P(\Delta)$, thereby artificially increasing the significance of subclustering and vice versa. 
\cite{Pinkney1996} highlights in their comparison of substructure tests how the sensitivity of the $\Delta$ test is affected measurably by the projection angle of the member galaxies, this can lead to a potential loss of genuine merging systems from our sub-sample when their velocities run along $0\degree$ or $90\degree$. 
One way to potentially alleviate this could be the introduction of more spatial parameters. 
For example, the \textit{Lee Three-Dimensional Statistic} adapted by \cite{Fitchett1987}, took into consideration angles derived from the projected space and velocity. 
This test can help to eliminate any potential false positive with its ability to be insensitive to genuine non-merging systems \citep{Pinkney1996}. 

There are also methods for testing dynamical activity that involve measuring the Gaussianity of the velocity distributions, such as the \textquoteleft Hellinger Distance' measuring the distance between a set of observational and theoretical distributions (see \citealt{Ribeiro2013}; \citealt{Carvalho2017}). 
Other novel approaches, such as one presented by \cite{Schwinn2018}, test to see whether 2D mass maps can be used to find mass peaks using wavelet transform coefficients.
Highlighting discrepancies between definitions of substructure.
In contrast, tried and tested methods are evaluated by \cite{Hou2009}, comparing different approaches to analysing the dynamical complexity to groups of galaxies. 
The authors find a $\chi^{2}$ goodness-of-fit is not best suited for determining a transition away from a Gaussian distribution of velocities. 
The principles upon which the $\Delta$ test is built upon is a frequentist $\chi^{2}$, which may indicate there is some form of decoupling in the link between sub-structuring and dynamical activity.
This apparent decoupling is most likely a result of the limitations of using a singular technique to define if a merging system of cluster galaxies is present, as the $\Delta$ test is only sensitive to average deviations from observed line-of-sight velocities.
This is a problem that extends to the VDPs, since they rely on a weighted grouping of objects in velocity-space with a moving Gaussian window function.
Therefore, care has to be taken when classifying a galaxy cluster as merging or non-merging based on using the methodology of \cite{Bergond2006} and \cite{Hou2009}.  
Despite these caveats, the nature of determining substructure with classical statistical testing is simple, sensitive and allows for fast computation on determining our sub-samples. 
However, there is room to consider how one can accurately define a cluster to be merging or not based solely on limiting velocity-space tests for substructure/grouping of galaxies.
For example, there are relic mergers with non-thermal emissions that represent an afterglow of a merging event, or, represent a pre-merging environment as a result from the interactions between intra-cluster media (e.g. \citealt{Giovannini2009}; \citealt{Bulbul2016}).
These environments would be insensitive to our traditional statistical testing for substructure due to its constrained application on using the clustering of galaxies as the sole proxy for a merging system. 
Utilising other parts of the spectrum highlight strong interactions between particles of the ICM, or, of two interacting ICMs from two initially independent systems, and the lack of a comprehensive study can call into question how we best define what is and is not a merging cluster.

The VDPs produced here could potentially mask any further variability within the kinematics that would otherwise be visible on a smaller scale \textquoteleft window width'. 
It is apparent from this work there is some form of sub-layer to the profiles that inhibit a clearer picture being formed in the dynamical nature of galaxies with differing properties. 
It is a notable possibility that, within some clusters, there is still an inclusion of interloper field galaxies towards $\sim2.5$ $r_{200}$ that distort our final view on the key drivers of these seemingly interacting galaxy sub-populations. 
The differing merging and non-merging sample sizes present problems of their own that lead to biasing the final stacked VDPs.
For example the smoothing kernel, along with the chosen width of the kernel, used will cause a decrease in the sensitivity in how the VDPs respond to substructuring.
This problem continues with the stacking procedures, which decrease the sensitivity to the presence of mergers due to each cluster possessing unique environments with different position angles and separations.
This problem is further extended when clusters possess limited numbers due to spectroscopic limitations of the survey in the MPA-JHU data. 
Therefore, unless there is a significant number of galaxies inputted to the calculation of a VDP, the risk of spurious features appearing is still a powerful one.  
In some cases this is purely a limitation of the data available from marrying the MPA-JHU with DR8 photometry or GZ2 morphologies, in others, an indicator to the limitations in using VDPs as a tool to present the dynamical overview of galaxy clusters.

\section{Summary}
\label{sec:conc}

In this work we have produced a base line cluster galaxy membership that marries the MPA-JHU DR8 archival data with the BAX limits of $(3 < L_{X} < 30)\times10^{44}$ ergs s$^{-1}$ and $0.0 < z < 0.1$, which is complete at log$_{10}(M_\ast)>10.1$.
The sample of galaxy clusters are sub-categorised into a merging or non-merging samples of galaxy clusters depending on the outcome of the \cite{Dressler1988} test.
Stacks of VDPs are computed for differing galactic parameters in order to determine what drives the shape of the VDP.

\noindent
The key results are summarised as follows:

\begin{itemize}[leftmargin=*] 

\item In common with previous literature, our merging cluster sample demonstrates a steeply rising VDP. The bulk of this rise happens at $\sim1.5$ $r_{200}$. On the other hand, non-merging clusters generally exhibit a declining-to-flat VDP.

\item In merging systems, a mix of red and blue elliptical galaxies appear to be driving the rising VDP at these radii. This may be the result of pre-processing within galaxy groups.

\item Non-merging systems commonly display little variation in kinematics throughout their VDPs, however, there are consistently higher $\sigma_{P}(R)$ values from the VDPs associated with a younger population of galaxies.

\item Spiral galaxy VDPs in merging systems present a dichotomy in their dispersion of velocities, with the blue spiral galaxies possessing a high velocity dispersion that is indicative to an infalling sub-population of field galaxies.
	
\item The global VDP of an individual cluster must be treated with care since a rising or falling VDP may be driven by a subpopulation of the cluster members.    

\end{itemize}

\section*{Acknowledgements}

We extend our thanks to the referee for their valued time in offering thorough feedback and suggestions to this work. Providing a valuable second opinion.

\noindent
We would like to thank Bill Harris for his assistance and advice in the confirmation of the weighted kernel formulation used within this work.

\noindent
LEB wishes to thank Yjan Gordon for his continued help and advice throughout the formulation of this work.

\noindent
We acknowledge the support of STFC through the University of Hull\textquotesingle s Consolidated Grant ST/R000840/1.

\noindent
This research made use of Astropy, a community-developed core Python package for Astronomy (\citealt{Collaboration2013}; \citealt{Collaboration2018}).

\noindent
This research has made use of the X-Rays Clusters Database (BAX)
which is operated by the Laboratoire d'Astrophysique de Tarbes-Toulouse (LATT),
under contract with the Centre National d'Etudes Spatiales (CNES) 

\noindent
This research has  made use of the \textquotedblleft K-corrections calculator'' service available at \url{http://kcor.sai.msu.ru/}

\noindent
Funding for SDSS-III has been provided by the Alfred P. Sloan Foundation, the Participating Institutions, the National Science Foundation, and the U.S. Department of Energy Office of Science. The SDSS-III web site is \url{http://www.sdss3.org/}.
SDSS-III is managed by the Astrophysical Research Consortium for the Participating Institutions of the SDSS-III Collaboration including the University of Arizona, the Brazilian Participation Group, Brookhaven National Laboratory, Carnegie Mellon University, University of Florida, the French Participation Group, the German Participation Group, Harvard University, the Instituto de Astrofisica de Canarias, the Michigan State/Notre Dame/JINA Participation Group, Johns Hopkins University, Lawrence Berkeley National Laboratory, Max Planck Institute for Astrophysics, Max Planck Institute for Extraterrestrial Physics, New Mexico State University, New York University, Ohio State University, Pennsylvania State University, University of Portsmouth, Princeton University, the Spanish Participation Group, University of Tokyo, University of Utah, Vanderbilt University, University of Virginia, University of Washington, and Yale University.





\bibliographystyle{mnras}
\bibliography{vdp} 

\bsp	
\label{lastpage}
\end{document}